\documentclass[onecolumn,a4paper,12pt,oneside]{article} 
\usepackage[top=2.5cm,left=2.5cm,right=2.5cm,bottom=3cm]{geometry}    
\usepackage[nostamp]{draftwatermark} 
\SetWatermarkText{\bf DRAFT}
\SetWatermarkAngle{45}
\SetWatermarkScale{4}
\SetWatermarkLightness{0.85}     
\usepackage[parfill]{parskip}    
\usepackage{graphicx}
\usepackage{amssymb}
\usepackage{amsmath}
\usepackage{amsfonts}
\usepackage{amsthm}
\usepackage{epstopdf}
\usepackage[usenames,dvipsnames,svgnames,table]{xcolor}
\usepackage{tikz}
\usepackage{array}
\usepackage[T1]{fontenc}
\usepackage[utf8]{inputenc}
\usepackage{booktabs} 
\usepackage{array} 
\usepackage{paralist} 
\usepackage{listings} 
\usepackage{verbatim} 
\usepackage{subfig} 
\usepackage[colorlinks=true]{hyperref}
\hypersetup{colorlinks=true, linktocpage=true, linkcolor=Blue, citecolor=Green,
hypertexnames=true, urlcolor=MidnightBlue,pdftex}
\usepackage[english]{babel}
\usepackage{wrapfig}
\usepackage{multirow}
\usepackage{quoting}
\usepackage{rotating}
\quotingsetup{font=normalsize}
\theoremstyle{plain}

\usepackage{bm}
\usepackage{abstract}
\usepackage{cite}
\usepackage{float}
\usepackage{textcomp} 	
\usepackage{tabularx} 
\usepackage{caption}
\usepackage{authblk} 
\usepackage{eso-pic} 
\usepackage{enumitem} 
\usepackage{lineno} 
\providecommand{\keywords}[1]{\textbf{{Key words: }} #1} 

\newcommand{\be}{\begin{equation}}
\newcommand{\ee}{\end{equation}}
\newcommand{\bsp}{\begin{split}}
\newcommand{\esp}{\end{split}}
\newcommand{\bsig}{\boldsymbol{\sigma}}

\renewcommand{\Phi}{\varPhi}

\newcommand{\eps}{\varepsilon}
\newcommand{\E}{\mathbb{E}}

\renewcommand{\Theta}{\varTheta}
\renewcommand{\Psi}{\varPsi}
\renewcommand{\Sigma}{\varSigma}

\renewcommand{\Delta}{\varDelta}
\renewcommand{\phi}{\varphi}
\renewcommand{\psi}{\varPsi}

\title{{\Large{\bf Elastic bounds for anisotropic layers}}\medskip\\}

\author{\begin{small}Paolo Vannucci\\ 
LMV - Laboratoire de Mathématiques de Versailles, UMR8100\\Université Paris-Saclay - UVSQ\\\href{mailto:paolo.vannucci@uvsq.fr}{paolo.vannucci@uvsq.fr}\end{small}\medskip\\

\small{ Preprint of : Bounds of the technical constants for two-dimensional anisotropic elasticity.\\Proc. Royal Society A, 20230662. \href{https://doi.org/10.1098/rspa.2023.0662}{https://doi.org/10.1098/rspa.2023.0662}}}

\date{}
\begin{document}
\maketitle

\hrule
\begin{abstract}
The complete set of bounds for the technical constants of an elastic layer, plate or laminate is given. The bounds are valid in general, also for completely anisotropic bodies. They are obtained transforming the polar bounds previously found. These bounds complete the knowledge of classical elasticity at least in the two-dimensional case and are useful in several situations, e.g., for determining the correct feasibility domain in design problems or as necessary conditions for accepting  the results of laboratory tests on anisotropic layers.

\end{abstract}
\medskip
\hrule
\bigskip

\section{Introduction}
The existence and determination of bounds for the  moduli of a material is a well known topic in the theory of elasticity. As well known, such bounds are the necessary result of the physical condition imposing the positiveness of the work done by the applied forces on an elastic body for deforming this one.

 This problem  can be solved using either purely mathematical, cf. \cite{hohn58}, or more directly mechanically inspired approaches, cf. \cite{Love}. For isotropic materials, this leads to the well known bounds for either the Lamé's parameters 
 \be
 3\lambda+2\mu>0,\ \ \mu>0,
 \ee
 or for the more commonly used technical parameters, i.e. the Young's modulus and the Poisson's ratio,
 \be
 E>0,\ \ -1<\nu<\frac{1}{2}.
 \ee

For anisotropic materials, a, basically mathematic, approach, based upon a somewhat forgotten theorem, cf. \cite{hohn58}, page 340, allows to give a general form to the bounds of the components of the stiffness (or alternatively of the compliance)  matrix $[C]$,
\be
\{\sigma\}=[C]\{\eps\},
\ee
that traduces, through the Kelvin's notation, \cite{kelvin,kelvin1}, the components of the fourth-order elastic tensor $\E$ into the components of a $6\times6$ symmetric matrix, 	according to the rule
\be
[C]\hspace{-1mm}=\hspace{-1mm}\begin{footnotesize}\left[\hspace{-1mm}\begin{array}{llllll}
C_{11}=\E_{1111}&C_{12}=\E_{1122}&C_{13}=\E_{1133}&C_{14}=\sqrt{2}\E_{1123}&C_{15}=\sqrt{2}\E_{1131}&C_{16}=\sqrt{2}\E_{1112}\\
                            &C_{22}=\E_{2222}&C_{23}=\E_{2233}&C_{24}=\sqrt{2}\E_{2223}&C_{25}=\sqrt{2}\E_{2231}&C_{26}=\sqrt{2}\E_{2212}\\
                            &                             &C_{33}=\E_{3333}&C_{34}=\sqrt{2}\E_{3323}&C_{35}=\sqrt{2}\E_{3331}&C_{36}=\sqrt{2}\E_{3312}\\
                            &                             &                             &C_{44}=2\E_{2323}         &C_{45}=2\E_{2331}         &C_{46}=2\E_{2312}\\
                            &                             &                             &                                        &C_{55}=2\E_{3131}         &C_{56}=2\E_{3112}\\
                            &                             &                             &                                        &                                        &C_{66}=2\E_{1212}
\end{array}\hspace{-1mm}\right]\end{footnotesize}\hspace{-1mm}.
\ee
By this theorem, the positiveness of $[C]$ is get imposing that the six leading principal minors of $[C]$ are all positive. This approach has the advantage of clearly fixing the number, six, of bounds to be written in the most general case; in the presence of some material symmetry, these bounds are less than six and can be written in an explicit form, cf. \cite{vannucci_libro}. 
 
However, and rather surprisingly, for three-dimensional anisotropic bodies the determination of the bounds for the technical moduli  is still an open problem, only partial results are known in the literature. In particular, cf. \cite{Lekhnitskii} or \cite{jones}, the following conditions are normally  given in the literature:
\be
\begin{array}{c}
\forall i,j\in\{1,2,3\},\ E_i>0,\ \ G_{ij}>0,\medskip\\
\dfrac{1-2\nu_{12}}{E_1}+\dfrac{1-2\nu_{23}}{E_2}+\dfrac{1-2\nu_{31}}{E_3}>0,\\
\end{array}
\ee
Another, rougher, bound is given by Lekhnitskii, \cite{Lekhnitskii},
\be
\nu_{12}+\nu_{23}+\nu_{31}<\frac{3}{2}.
\ee
These are the only bounds valid for any elastic body, regardless from its syngony, i.e. they are valid also for triclinic materials. Some other bounds are known but uniquely for bodies that are at least orthotropic and with the moduli measured in a reference frame whose  axes correspond with some equivalent directions for the material:
\be
\begin{array}{c}
1-\nu_{ij}\nu_{ji}>0\ \ \forall i,j\in\{1,2,3\},\medskip\\
1-\nu_{12}\nu_{21}-\nu_{23}\nu_{32}-\nu_{31}\nu_{13}-2\nu_{32}\nu_{21}\nu_{13}>0,
\end{array}
\ee
conditions that can be transformed, respectively, to
\be
\label{eqs:bornes3D}
\begin{array}{c}
|\nu_{ij}|<\sqrt{\dfrac{E_i}{E_j}},\medskip\\
\nu_{32}\nu_{21}\nu_{13}<\dfrac{1}{2}\left(1-\nu_{32}^2\dfrac{E_2}{E_3}-\nu_{21}^2\dfrac{E_1}{E_2}-\nu_{13}^2\dfrac{E_3}{E_1}\right)<\dfrac{1}{2}.
\end{array}
\ee
All these bounds on the technical constants are 
found imposing the positiveness of the strain energy for peculiar stress fields, e.g.  pure extension or shear.

Unlike in the general three-dimensional case, in planar elasticity the problem has been completely solved in a particular set of elastic moduli, the so-called {\it polar parameters}. However, the corresponding of the polar bounds for the technical constants have never been given. This is the topic of this paper, organized as follows: in the next Section, the polar bounds are recalled, then the bounds for the technical constants in planar elasticity are obtained and finally some particular cases discussed.

\section{Polar bounds}
The polar formalism was introduced in 1979 by G. Verchery, \cite{Verchery79}; a complete account of the method can be found in \cite{Meccanica05,vannucci_libro}, here only some elements of this method, necessary to the developments, are recalled.

By the polar method, the Cartesian components at a direction $\theta$ of a matrix, e.g. $[C]$, representing, in the Kelvin notation, a plane elastic tensor, e.g. $\E$,  are expressed as
\begin{equation}
\label{eq:mohr4}
{\begin{array}{l}
{C_{11}{(}\theta{)}={T}_{0}{+}{2}{T}_{1}{+}{R}_{0}\cos{4}\left({{{\varPhi}}_{0}{-}\theta}\right){+}{4}{R}_{1}\cos{2}\left({{{\varPhi}}_{1}{-}\theta}\right)},\medskip\\
{C_{12}{(}\theta{)}={-}{T}_{0}{+}{2}{T}_{1}{-}{R}_{0}\cos{4}\left({{{\varPhi}}_{0}{-}\theta}\right)},\medskip\\
C_{16}(\theta)=\sqrt{2}\left[R_0\sin4\left(\varPhi_0-\theta\right)+2R_1\sin2\left(\varPhi_1-\theta\right)\right],\medskip\\
{C_{22}{(}\theta{)}={T}_{0}{+}{2}{T}_{1}{+}{R}_{0}\cos{4}\left({{{\varPhi}}_{0}{-}\theta}\right){-}{4}{R}_{1}\cos{2}\left({{{\varPhi}}_{1}{-}\theta}\right)},\medskip\\
{C_{26}{(}\theta{)}=\sqrt{2}\left[{-}{R}_{0}\sin{4}\left({{{\varPhi}}_{0}{-}\theta}\right){+}{2}{R}_{1}\sin{2}\left({{{\varPhi}}_{1}{-}\theta}\right)\right]},\medskip\\
{C_{66}{(}\theta{)}=2\left[{T}_{0}{-}{R}_{0}\cos{4}\left({{{\varPhi}}_{0}{-}\theta}\right)\right]}.
\end{array}}
\end{equation}
The moduli $T_0,T_1,R_0,R_1$ as well as the difference of the angles $\Phi_0-\Phi_1$ are tensor invariants.  The choice of one of the two polar angles fixes the frame; usually $\Phi_1=0$. 

The converse of the previous equations are
\be
\label{eq:conversepolar}
\begin{array}{l}
T_0=\dfrac{1}{8}(C_{11}-2C_{12}+2C_{66}+C_{22}),\medskip\\
T_1=\dfrac{1}{8}(C_{11}+2C_{12}+C_{22}),\medskip\\
R_0=\dfrac{1}{8}\sqrt{(C_{11}-2C_{12}-2C_{66}+C_{22})^2+8(C_{16}-C_{26})^2},\medskip\\
R_1=\dfrac{1}{8}\sqrt{(C_{11}-C_{22})^2+2(C_{16}+C_{26})^2},\medskip\\
\tan4\Phi_0=\dfrac{2\sqrt{2}(C_{16}-C_{26})}{C_{11}-2C_{12}-2C_{66}+C_{22}},\medskip\\
\tan2\Phi_1=\dfrac{2\sqrt{2}(C_{16}+C_{26})}{C_{11}-C_{22}}.
\end{array}
\ee

The elastic symmetries are determined by the following conditions on the invariants:
\begin{itemize}
\item ordinary orthotropy: $\Phi_0-\Phi_1=K\dfrac{\pi}{4}, \ K\in\{0,1\}$;
\item $R_0$-orthotropy: $R_0=0$, \cite{vannucci02joe};
\item square symmetry: ${R_1=0}$;
\item isotropy: $R_0=R_1=0$.
\end{itemize}
So we see that $T_0$ and $T_1$ are the {\it isotropy invariants}, while $R_0,R_1$ and $\Phi_0-\Phi_1$ are the {\it anisotropy invariants}. 
The above relations are valid for any matrix of the elastic type, hence  for  the compliance matrix $[S]=[C]^{-1}$ too; we will indicate by $t_0,t_1,r_0,r_1,\phi_0$ and $\phi_1$ the polar parameters of $[S]$.

For  the components of the vector, say $\{L\}$, representing in the Kelvin formalism a second-rank symmetric tensor {\bf L},  the polar formalism gives
\be
\label{eq:mohr}
\begin{split}
&L_{1}(\theta)=T+R\cos2(\varPhi-\theta),\medskip\\
&L_{2}(\theta)=T-R\cos2(\varPhi-\theta),\medskip\\
&L_{6}(\theta)=\sqrt{2}R\sin2(\varPhi-\theta),
\end{split}
\ee
with $T, R$ two invariants, representing respectively the {\it isotropic} and the {\it anisotropic} phases of {\bf L}; $\Phi$ is an angle determined by the choice of the frame.

The polar bounds can be found in the following way, \cite{vannucci_libro}: the strain energy density per unit volume,
\be
V_\eps=\frac{1}{2}\{\sigma\}^\top\{\eps\}=\frac{1}{2}\{\eps\}^\top[C]\{\eps\},
\ee
can be written using the polar components of $[C]$ and $\{\eps\}$ through eqs.  (\ref{eq:mohr4}) and (\ref{eq:mohr}):
\be
V_\eps=4T_1 \ t^2+8R_1\cos2(\Phi_1-\phi)r\ t+2[T_0+R_0\cos4(\Phi_0-\phi)]r^2,
\ee
where $t,r$ and $\phi$ are the polar parameters of $\{\eps\}$. This quantity can be rewritten as
\be
V_\eps=\{r,t\}\left[
\begin{array}{cc}
2[T_0+R_0\cos4(\Phi_0-\phi)]&4R_1\cos2(\Phi_1-\phi)\\
4R_1\cos2(\Phi_1-\phi)&4T_1
\end{array}
\right]
\left\{
\begin{array}{c}r\\t\end{array}\right\}.
\ee
The positivity of $V_\eps\ \forall\{r,t\}$, stating  the physical condition of a positive work done by the applied forces, is ensured if and only if the matrix in the previous equation is positive definite; by the already mentioned theorem on the leading principal minors, \cite{hohn58}, this happens if and only if the following two conditions are satisfied:
\be
\label{eq:bornes1}
\left\{\begin{array}{l}
T_0+R_0\cos4(\Phi_0-\phi)\medskip\\
T_1[T_0+R_0\cos4(\Phi_0-\phi)]-2R_1^2\cos^22(\Phi_1-\phi)>0
\end{array}\ \ \ \ \forall\phi.\right.
\ee
These two conditions can be transformed to three other inequalities and it can be proved that one of them is redundant (the complete, technical, proof is omitted here, the reader is addressed to \cite{vannucci15ijss} or to \cite{vannucci_libro}). In the end, one gets the bounds
\be
\label{eq:bornes2}
\begin{array}{l}
T_0-R_0>0,\medskip\\
T_1\left(T_0^2-R_0^2\right)-2R_1^2[T_0-R_0\cos4(\Phi_0-\Phi_1)]>0.
\end{array}
\ee
To remark that, being moduli of complex numbers, $R_0\ge0,R_1\ge0$, which necessarily implies, through eqs. (\ref{eq:bornes1}) and (\ref{eq:bornes2})$_1, T_0>0,T_1>0$.

The above bounds are general, i.e. valid for any type of layer, and are written in terms of polar invariant, so they are {\it intrinsic bounds}, i.e. frame independent. They can be applied as well to the polar parameters of $[C]$ or of $[S]$:
\be
\label{eq:bornes3}
\begin{array}{l}
t_0-r_0>0,\medskip\\
t_1\left(t_0^2-r_0^2\right)-2r_1^2[t_0-r_0\cos4(\phi_0-\phi_1)]>0.
\end{array}
\ee
These last bounds can be obtained following the same procedure but starting from the stress energy
\be
V_\sigma=\frac{1}{2}\{\sigma\}^\top[S]\{\sigma\}.
\ee
Of course, the same bounds can be written also for any other elastic-type planar tensor, like, for instance, the extension and bending stiffness, and compliance, tensors of laminates, see e.g. \cite{jones,gay14,vannucci_libro}.

\section{The bounds for the technical constants}
In order to find the bounds for the technical constants, first, eq. (\ref{eq:conversepolar}) is written for the components of $[S]$:
\be
\label{eq:conversecompl}
\begin{array}{l}
t_0=\dfrac{1}{8}(S_{11}-2S_{12}+2S_{66}+S_{22}),\medskip\\
t_1=\dfrac{1}{8}(S_{11}+2S_{12}+S_{22}),\medskip\\
r_0=\dfrac{1}{8}\sqrt{(S_{11}-2S_{12}-2S_{66}+S_{22})^2+8(S_{16}-S_{26})^2},\medskip\\
r_1=\dfrac{1}{8}\sqrt{(S_{11}-S_{22})^2+2(S_{16}+S_{26})^2},\medskip\\
\tan4\phi_0=\dfrac{2\sqrt{2}(S_{16}-S_{26})}{S_{11}-2S_{12}-2S_{66}+S_{22}},\medskip\\
\tan2\phi_1=\dfrac{2\sqrt{2}(S_{16}+S_{26})}{S_{11}-S_{22}}.
\end{array}
\ee
Then, this result is used into eq. (\ref{eq:bornes3}): the first condition becomes
\be
\dfrac{1}{8}(S_{11}-2S_{12}+2S_{66}+S_{22})>\dfrac{1}{8}\sqrt{(S_{11}-2S_{12}-2S_{66}+S_{22})^2+8(S_{16}-S_{26})^2},
\ee
which gives the two conditions (a third one, stating that the argument of the square root at the second member must be positive, is redundant)
\be
\begin{array}{l}
S_{11}-2S_{12}+2S_{66}+S_{22}>0,\medskip\\
S_{66}(S_{11}-2S_{12}+S_{22})>(S_{16}-S_{26})^2.
\end{array}
\ee
To transform eq. (\ref{eq:bornes3})$_2$, it is worth to introduce the following polar invariant, \cite{vannucci_libro}, page 145:
\be
c_1=8r_1^2r_0\cos4(\phi_0-\phi_1)
\ee
which gives
\be
\label{eq:r0c1}
r_0\cos4(\phi_0-\phi_1)=\frac{c_1}{8r_1^2}.
\ee
The Cartesian expression of $c_1$ is known:
\be
\begin{split}
c_1=&\frac{1}{64}\left[(S_{11}-S_{22})^2-2(S_{16}+S_{26})^2\right](S_{11}-2S_{12}-2S_{66}+S_{22})+\\
&+\frac{1}{8}(S_{11}-S_{22})\left(S_{16}^2+S_{26}^2\right).
\end{split}
\ee
Finally, eq. (\ref{eq:bornes3})$_2$ becomes first
\be
t_1\left(t_0^2-r_0^2\right)>2r_1^2-\frac{c_1}{4},
\ee
then, after some standard passages,
\be
2S_{12}S_{16}S_{26}+S_{11}S_{22}S_{66}-S_{22}S_{16}^2-S_{11}S_{26}^2-S_{66}S_{12}^2>0.
\ee
To remark that actually
\be
2S_{12}S_{16}S_{26}+S_{11}S_{22}S_{66}-S_{22}S_{16}^2-S_{11}S_{26}^2-S_{66}S_{12}^2=\det[S].
\ee
We get hence the three bounds for the Cartesian components of $[S]$
\be
\label{eq:bornes4}
\begin{array}{l}
S_{11}-2S_{12}+2S_{66}+S_{22}>0,\medskip\\
S_{66}(S_{11}-2S_{12}+S_{22})>(S_{16}-S_{26})^2,\medskip\\
2S_{12}S_{16}S_{26}+S_{11}S_{22}S_{66}-S_{22}S_{16}^2-S_{11}S_{26}^2-S_{66}S_{12}^2>0.
\end{array}
\ee
It is worth noting that this is not the only set of independent bounds that can be found for the $S_{ij}$s: applying to $[S]$ the already cited theorem on the leading principal minors, one should get three other bounds:
\be
\label{eq:bornes5}
\begin{array}{l}
S_{11}>0,\medskip\\
S_{11}S_{22}-S_{12}^2>0,\medskip\\
2S_{12}S_{16}S_{26}+S_{11}S_{22}S_{66}-S_{22}S_{16}^2-S_{11}S_{26}^2-S_{66}S_{12}^2>0.
\end{array}
\ee
The main difference between the last two sets of bounds is that  conditions (\ref{eq:bornes4}) uses exclusively invariant quantities, which is not the case for conditions (\ref{eq:bornes5}). Of course, imposing the positivity of the strain energy one can get similar relations for the $C_{ij}$s.

The passage to the technical constants can be made recalling that, by definition,
\be
\label{eq:sijtc}
\begin{array}{lll}
S_{11}=\dfrac{1}{E_1},& S_{12}=-\dfrac{\nu_{12}}{E_1}, & S_{16}=\dfrac{\eta_{12,1}}{\sqrt{2}E_1},\medskip\\
S_{22}=\dfrac{1}{E_2},& S_{66}=\dfrac{1}{2G_{12}},& S_{26}=\dfrac{\eta_{12,2}}{\sqrt{2}E_2},
\end{array}
\ee
with $E_1,E_2$ the Young's moduli in the directions of the two frame axes, $G_{12}$ the in-plane shear modulus, $\nu_{12}$ the in-plane Poisson's ratio and $\eta_{12,1},\eta_{12,2}$ two coefficients of mutual influence of the second type, \cite{jones, vannucci_libro}.
Alternatively, the coefficients of mutual influence of the first type $\eta_{k,ij}$ can be used, they are linked to the $\eta_{ij,k}$s by the reciprocity relations
\be
\frac{\eta_{ij,k}}{E_k}=\frac{\eta_{k,ij}}{G_{ij}},\ \ i,j,k\in\{1,2\},\ i\neq j.
\ee
Injecting eq. (\ref{eq:sijtc}) into eq. (\ref{eq:bornes4}) gives finally the bounds
\be
\label{eq:bornes6}
\begin{array}{l}
\dfrac{1+2\nu_{12}}{E_1}+\dfrac{1}{E_2}+\dfrac{1}{G_{12}}>0,\medskip\\
\dfrac{1}{E_1^2E_2^2G_{12}}\left\{E_1E_2[E_1+E_2(1+2\nu_{12})]-G_{12}(E_2\eta_{12,1}-E_1\eta_{12,2})^2\right\}>0,\medskip\\
\dfrac{1}{E_1^2E_2^2G_{12}}\left\{E_1(E_2-G_{12}\eta_{12,2}^2)-E_2[E_2\nu_{12}^2+G_{12}\eta_{12,1}(\eta_{12,1}+2\eta_{12,2}\nu_{12})]\right\}>0.
\end{array}
\ee
These three bounds are a set of necessary and sufficient conditions for the elastic energy density is positive for each stress/strain state. They use invariant quantities, which is undoubtedly an advantage in anisotropic elasticity.

If, in place of eq. (\ref{eq:bornes4}) we inject eq. (\ref{eq:sijtc}) into eq. (\ref{eq:bornes5}) we will get
\be
\label{eq:bornes7}
\begin{array}{l}
\dfrac{1}{E_1}>0,\medskip\\
\dfrac{1}{E_1E_2}-\dfrac{\nu_{12}^2}{E_1^2}>0,\medskip\\
\dfrac{1}{E_1^2E_2^2G_{12}}\left\{E_1(E_2-G_{12}\eta_{12,2}^2)-E_2[E_2\nu_{12}^2+G_{12}\eta_{12,1}(\eta_{12,1}+2\eta_{12,2}\nu_{12})]\right\}>0.
\end{array}
\ee
The third condition remains the same, while the two first ones can be rewritten as
\be
E_1>0,\ \ |\nu_{12}|<\sqrt{\frac{E_1}{E_2}}.
\ee
The first one is the well-known condition of positiveness of the Young's moduli, while the second one corresponds to the bound (\ref{eqs:bornes3D})$_1$. Also, these two bounds, or alternatively the fact that $E_1>0$ for any direction, imply that it is $E_2>0$ too. Finally, also $G_{12}>0$, which can be  proved, classically, with the mechanical  procedure imposing a pure shear stress state or, using a purely mathematical approach, still using the theorem on leading principal minors once reordered the vector $\{\sigma\}$ representing, in the Kelvin notation, the stress tensor $\bsig$ as $\{\sigma\}=\{\sigma_6,\sigma_1,\sigma_2\}^\top$.

Though eq. (\ref{eq:bornes6}) are a minimal set of necessary and sufficient conditions for ensuring the positivity of the elastic energy, it is perhaps better, from a practical point of view, to dispose of more direct bounds, concerning, if possible, some quantities easy to be measured in laboratory tests. This can be done transforming eq. (\ref{eq:bornes6})  and taking into account for the positivity of the Young's and shear moduli. Some short passages lead to the set of conditions
\be
\label{eq:bornes8}
\begin{array}{l}
E_1>0,\medskip\\
E_2>0,\medskip\\
G_{12}>0,\medskip\\
E_1E_2+G_{12}[E_1+E_2(1+2\nu_{12})]>0,\medskip\\
E_1E_2[E_1+E_2(1+2\nu_{12})]-G_{12}(E_2\eta_{12,1}-E_1\eta_{12,2})^2>0,\medskip\\
E_1(E_2-G_{12}\eta_{12,2}^2)-E_2[E_2\nu_{12}^2+G_{12}\eta_{12,1}(\eta_{12,1}+2\eta_{12,2}\nu_{12})]>0.
\end{array}
\ee
The above bounds, however, do not use invariant quantities. Though formed by a redundant number of bounds, this set of conditions is perhaps more interesting for practical applications, namely for checking the results of laboratory tests used for characterizing an anisotropic layer or  plate. To remark that in this set of bounds the coefficients of mutual influence enter the problem; in all the bounds known in the literature, these coefficients were absent.

\section{Bounds for layers with material symmetries}
Let us consider now how bounds (\ref{eq:bornes8}) change where some kind of material symmetry is present.

{\bf Isotropy.} In this case $r_0=R_0=r_1=R_1=0$, so the polar bounds reduce simply to 
\be
t_0>0,\ \ t_1>0,
\ee
while for the Cartesian components we have
\be
S_{11}=S_{22}=t_0+2t_1,\ \ S_{12}=-t_0+2t_1,\ \ S_{66}=2t_0=S_{11}-S_{12},\ \ S_{16}=S_{26}=0,
\ee
which gives
\be
t_0=\frac{S_{11}-S_{12}}{4}, \ \ t_1=\frac{S_{11}+S_{12}}{4},
\ee
so the Cartesian bounds are simply
\be
S_{11}-S_{12}>0,\ \ S_{11}+S_{12}>0.
\ee
Then, because for isotropy 
\be
E_1=E_2:=E,\ \ \nu_{12}:=\nu,\ \ G_{12}:=G=\dfrac{E}{2(1+\nu)},\ \ \eta_{12,1}=\eta_{12,2}=0,
\ee
bounds (\ref{eq:bornes8}) reduce to
\be
E>0,\ \ G>0,\ \ 2E^2>0,\ \ 2E^3(1+\nu)>0,\ \ E^2(1-\nu^2)>0,
\ee
which of course give the three well known bounds for isotropic planar bodies
\be
E>0,\ \ -1<\nu<1,
\ee
or 
\be
\lambda+\mu>0,\ \ \mu>0
\ee
with the Lamé's parameters.

{\bf Ordinary orthotropy.} 
The polar condition for orthotropy is 
\be
\phi_0-\phi_1=k\frac{\pi}{4},\ \ k\in\{0,1\}.
\ee
The discussion of the mechanical and mathematical differences between the two types of ordinary orthotropy, $k=0$ or $k=1$, can be found in \cite{Meccanica05,vannucci_libro}. However, for the orthotropic case,
\be
r_0\cos4(\phi_0-\phi_1)=(-1)^kr_0,
\ee
but eq. (\ref{eq:r0c1}) does not change, so bounds (\ref{eq:bornes8}) remain the same: there is no difference for the two ordinarily orthotropic cases. To remark that bounds (\ref{eq:bornes8}) are written in a generic frame, not necessarily in the orthotropic one. If the axes coincide with the orthotropy directions, then $\eta_{12,1}=\eta_{12,2}=0$ and bounds (\ref{eq:bornes8}) become
\be
\begin{array}{l}
E_1>0,\medskip\\
E_2>0,\medskip\\
G_{12}>0,\medskip\\
E_1+E_2(1+2\nu_{12})>0,\medskip\\
E_2(E_1-E_2\nu_{12}^2)>0,
\end{array}
\ee
the condition (\ref{eq:bornes8})$_4$ becoming now redundant.

{\bf Square-symmetry.} This special type of orthotropy is determined by the polar condition $r_1=R_1=0$. The polar bounds (\ref{eq:bornes3}) become
\be
\begin{array}{l}
t_0-r_0>0,\medskip\\
t_1(t_0^2-r_0^2)>0,
\end{array}
\ee
which, being $r_0>0$, reduce to
\be
t_0-r_0>0,\ \ t_1>0.
\ee
For the Cartesian components we have, for any direction,
\be
S_{11}=S_{22},\ \ S_{26}=-S_{16},
\ee
so conditions (\ref{eq:bornes4}) become
\be
\begin{array}{l}
S_{11}-S_{12}+S_{66}>0,\medskip\\
S_{66}(S_{11}-S_{12})>2S_{16}^2,\medskip\\
S_{11}+S_{12}>0.
\end{array}
\ee
For the technical constants it is
\be
E_1=E_2,\ \ \eta_{12,2}=-\eta_{12,1}\frac{E_2}{E_1},
\ee
so bounds (\ref{eq:bornes8}) become
\be
\begin{array}{l}
E_1>0,\medskip\\
G_{12}>0,\medskip\\
E_1+2G_{12}(1+\nu_{12})>0,\medskip\\
E_1(1+\nu_{12})-2G_{12}\eta_{12,1}^2>0,\medskip\\
1-\nu_{12}>0.
\end{array}
\ee
When written in one of the two symmetry frames, shifted of $\pi/4$, because $\eta_{12,1}=0$ we get
\be
\begin{array}{l}
E_1>0,\medskip\\
G_{12}>0,\medskip\\
E_1+2G_{12}(1+\nu_{12})>0,\medskip\\
-1<\nu_{12}<1.
\end{array}
\ee

{\bf $R_0$-orthotropy.} This special orthotropy is characterized by $R_0=0$, \cite{vannucci02joe}. However, unlike the case of square symmetry, this does not imply that $r_0=0$, but that
\be
r_0=\frac{r_1^2}{t_1},\ \ k=0,
\ee
so that bounds (\ref{eq:bornes3}) become
\be
\begin{array}{l}
t_0-\dfrac{r_1^2}{t_1}>0,\medskip\\
t_1\left(t_0+\dfrac{r_1^2}{t_1}\right)-2r_1^2>0.
\end{array}
\ee
Rather surprisingly, the bounds on the $S_{ij}$s, eq. (\ref{eq:bornes4}) reduce to only two, the third one being a product of the first two ones (this can be checked using eqs. (\ref{eq:conversecompl})$_{1,2,4}$ into the previous equations):
\be
\begin{array}{l}
S_{11}+2S_{12}+S_{22}>0,\medskip\\
2(S_{11}S_{22}-S_{12}^2)+S_{66}(S_{11}+2S_{12}+S_{22})-(S_{16}+S_{26})^2>0.
\end{array}
\ee
Injecting eq. (\ref{eq:sijtc}) into these conditions, the bounds for the technical constants are obtained as well:
\be
\begin{array}{l}
E_1>0,\medskip\\
E_2>0,\medskip\\
G_{12}>0,\medskip\\
E_1+E_2(1-2\nu_{12})>0,\medskip\\
E_1^2(E_2-G_{12}\eta_{12,2}^2)-E_2^2G_{12}(\eta_{12,1}^2+4\nu_{12}^2)+\medskip\\
+E_1E_2[E_2(1-2\nu_{12})+2G_{12}(2-\eta_{12,1}\eta_{12,2})]>0.
\end{array}
\ee

{\bf $r_0$-orthotropy.} This is the compliance dual of the previous case and it characterizes some special materials like paper,\cite{vannucci10joe}. In such a case, the polar bounds (\ref{eq:bornes3}) reduce to
\be
\label{eq:bornes9}
\begin{array}{l}
t_0>0,\medskip\\
t_0t_1-2r_1^2>0.
\end{array}
\ee
Concerning the components of $[S]$, 
\begin{equation}
\label{eq:Sr0}
{\begin{array}{l}
{S_{11}{(}\theta{)}={t}_{0}{+}{2}{t}_{1}{+}{4}{r}_{1}\cos{2}\left({{{\varphi}}_1-\theta}\right)},\medskip\\
{S_{12}{(}\theta{)}={-}{t}_{0}{+}{2}{t}_{1}},\medskip\\
S_{16}(\theta)=S_{26}{(}\theta{)}=2\sqrt{2}r_1\sin2\left(\varphi_1-\theta\right),\medskip\\
{S_{22}{(}\theta{)}={t}_{0}{+}{2}{t}_{1}{-}{4}{r}_{1}\cos{2}\left({{{\varphi}}_1-\theta}\right)},\medskip\\
{S_{66}{(}\theta{)}=2{t}_{0}}.
\end{array}}
\end{equation}
It is worth noting that $S_{12}$ and $S_{66}$ are isotropic and that from the previous equations it is immediate to get
\be
t_0=\frac{1}{2}S_{66},\ \ t_1=\frac{1}{2}\left(S_{12}+\frac{S_{66}}{2}\right),\ \ 2r_1^2=\frac{1}{32}\left[(S_{11}-S_{22})^2+8S_{16}^2\right],
\ee
so bounds (\ref{eq:bornes9}) become
\be
S_{66}>0,\ \ \ 4S_{66}(S_{66}+2S_{12})-(S_{11}-S_{22})^2-8S_{16}^2>0
\ee
which for the technical parameters gives
\be
\label{eq:bornes10}
E_1>0,\ \ E_2>0,\ \ G_{12}>0,\ \ E_1E_2^2(E_1-4\nu_{12}G_{12})-(E_2-E_1)^2G_{12}^2-4\eta_{12,1}^2E_2^2G_{12}^2>0.
\ee
A much simpler form of these bounds can be obtained for a frame that coincides with the orthotropy axes ($\phi_1-\theta=0$); in such a case, from eq. (\ref{eq:Sr0}) we get $S_{16}=0$ and
\be
t_0=\frac{1}{4G_{12}},\ \ t_1=\frac{1}{2}\left(\frac{1}{4G_{12}}-\frac{\nu_{12}}{E_1}\right),\ \ r_1=\frac{1}{4}\left(\frac{1+\nu_{12}}{E_1}-\frac{1}{2G_{12}}\right),
\ee
so that bounds (\ref{eq:bornes10}) become
\be
E_2>0,\ \ G_{12}>0,\ \ E_1>G_{12}(1+\nu_{12})^2.
\ee

\section{Conclusion}
The results found above fill the gap for what concerns the elastic bounds of the technical constants of an anisotropic layer or plate. What is important to point out is that they are written in a generic frame, so involving, and this is the first time, also the coefficients of mutual influence of the second type (through reciprocity relations, the use of the coefficients of the first type is straightforward). The interest of these bounds is first theoretical, then practical: the determination of the elastic parameters of a layer or of a plate is particularly difficult when the body is anisotropic. Disposing of bounds, serving as a necessary control for the acceptance of the experimental measures, is rather important in practice. The same can be said for correctly defining the feasibility domain in some design problem, e.g. concerning laminated plates made of composite, anisotropic layers.

\bibliographystyle{SageV} 

\bibliography{Biblio}   

\begin{thebibliography}{10}
\providecommand{\url}[1]{\texttt{#1}}
\providecommand{\urlprefix}{URL }
\expandafter\ifx\csname urlstyle\endcsname\relax
  \providecommand{\doi}[1]{DOI:\discretionary{}{}{}#1}\else
  \providecommand{\doi}{DOI:\discretionary{}{}{}\begingroup
  \urlstyle{rm}\Url}\fi
\providecommand{\eprint}[2][]{\url{#2}}

\bibitem{hohn58}
Hohn FE.
\newblock \emph{Elementary matrix algebra}.
\newblock New York, NY: MacMillan, 1958.

\bibitem{Love}
Love AEH.
\newblock \emph{A treatise on the mathematical theory of elasticity}.
\newblock New York, NY: Dover, 1944.

\bibitem{kelvin}
{Thomson W - Lord Kelvin}.
\newblock Elements of a mathematical theory of elasticity.
\newblock \emph{Philosophical Transations of the Royal Society} 1856; 146:
  481--498.

\bibitem{kelvin1}
{Thomson W - Lord Kelvin}.
\newblock Mathematical theory of elasticity.
\newblock \emph{Encyclopedia Britannica} 1878; 7: 819--825.

\bibitem{vannucci_libro}
Vannucci P.
\newblock \emph{Anisotropic elasticity}.
\newblock Berlin, Germany: Springer, 2018.

\bibitem{Lekhnitskii}
Lekhnitskii SG.
\newblock \emph{Theory of elasticity of an anisotropic elastic body}.
\newblock San Francisco, CA: English translation (1963) by {P. Fern},
  Holden-Day, 1950.

\bibitem{jones}
Jones RM.
\newblock \emph{Mechanics of composite materials. Second Edition}.
\newblock Philadelphia, PA: Taylor \& Francis, 1999.

\bibitem{Verchery79}
Verchery G.
\newblock {Les invariants des tenseurs d'ordre 4 du type de
  l'{\'e}lasticit{\'e}}.
\newblock In \emph{{Proc. of Colloque Euromech 115 (Villard-de-Lans, 1979):
  Comportement m{\'e}canique des mat{\'e}riaux anisotropes}}. Paris: Editions
  du CNRS, 1982, pp. 93--104.

\bibitem{Meccanica05}
Vannucci P.
\newblock Plane anisotropy by the polar method.
\newblock \emph{Meccanica} 2005; 40: 437--454.

\bibitem{vannucci02joe}
Vannucci P.
\newblock A special planar orthotropic material.
\newblock \emph{Journal of Elasticity} 2002; 67: 81--96.

\bibitem{vannucci15ijss}
Vannucci P and Desmorat B.
\newblock Analytical bounds for damage induced planar anisotropy.
\newblock \emph{International Journal of Solids and Structures} 2015; 60-61:
  96--106.

\bibitem{gay14}
Gay D.
\newblock \emph{{Composite Materials Design and Applications - Third Edition}}.
\newblock Boca Raton, FL: CRC Press, 2014.

\bibitem{vannucci10joe}
Vannucci P.
\newblock On special orthotropy of paper.
\newblock \emph{Journal of Elasticity} 2010; 99: 75--83.

\end{thebibliography}

\end{document}